\documentclass[12pt]{article}

\usepackage{scicite}

\usepackage{times}

\usepackage{gensymb}
\usepackage{mathptmx}
\usepackage{graphicx}
\usepackage[outdir=./]{epstopdf}
\usepackage[figurename=Fig.]{caption}
\usepackage{xcolor}
\usepackage{commath}
\usepackage{ulem}

\newenvironment{sciabstract}{%
\begin{quote} \bf}
{\end{quote}}

\title{Contactless Manipulation of Binary Droplets via Sensing of Localized Vapor Sources} 

\author
{Robert Malinowski$^{1}$, Ivan P. Parkin$^{1}$, Giorgio Volpe$^{1\ast}$ \\
\\
\normalsize{$^{1}$Department of Chemistry, University College London,}\\
\normalsize{20 Gordon Street, London WC1H 0AJ, United Kingdom}\\
\\
\normalsize{$^\ast$To whom correspondence should be addressed; E-mail: g.volpe@ucl.ac.uk.}
}

\date{}

\begin{document} 


\baselineskip24pt


\maketitle 


\begin{sciabstract}

Droplet motion on surfaces influences phenomena as diverse as microfluidic liquid handling, printing technology and micro-organism migration. Typically, droplet motion is achieved by inducing energy gradients on a substrate or flow on the droplet's free surface. Current configurations for droplet manipulation have, however, limited applicability as they rely on carefully tailored wettability gradients and/or bespoke substrates. Here we demonstrate the contactless long-range manipulation of binary droplets on pristine substrates due to the sensing of localized water vapor sources. We show with analytical considerations that the dissipative nature of the driving forces at play, induced by tiny asymmetries in surface tension gradients, is essential to capture the droplet's manipulation mechanism. We then demonstrate its versatility by printing, aligning and reacting materials controllably in space and time. 

\end{sciabstract}

\subsection*{Introduction}

Droplets moving on solid surfaces are at the heart of many phenomena of fundamental and applied interest in physics, biophysics, chemistry and materials science \cite{Lach2016}. Examples include ``tears of wine'' due to the Marangoni effect \cite{Nikolov2018}, rolling droplets on self-cleaning substrates \cite{Lu2015,Zhang2016}, microfluidic liquid handling \cite{Pollack2002,Holmes2015} as well as enhancing heat transfer \cite{Daniel2001}, sensing \cite{Cira2015}, and printing technologies \cite{Konvalina2015,Homede2016}. Even microorganisms, such as {\it Bacillus subtilis}, can collectively initiate motion of macroscopic water droplets, thus inducing the colony to move on surfaces \cite{Hennes2017}. Achieving real-time control over the directionality of moving droplets is therefore an important milestone towards harnessing them for applications in, e.g., printable materials \cite{Konvalina2015,Homede2016}, biological assays \cite{Srinivasan2004,Gulka2014}, and microreactors \cite{Yang2018}. 

The main impediment to trigger droplet motion on solid surfaces comes from the hysteresis of the contact angle that pins the droplet's edge to the underlying substrate \cite{Gao2006}. The generation of gradients of surface energy on this substrate is a widespread strategy to overcome the pinning of the contact line and achieve motion \cite{Chaudhury1992,Lee2000,Ichimura2000,Daniel2001,Liu2016,Jiang2016,Pollack2002,Tian2016}. Alternatively, imbalances of surface tension can be directly induced on the droplet's free surface, thus inducing flows within the droplet that ultimately lead to its motion \cite{Cira2015,Gao2018,Irajizad2017}. An emblematic, recent example of the latter mechanism is that of binary droplets of food colouring self-sensing their evaporation profile \cite{Cira2015,Wen2019,Sadafi2019}. Due to their widespread use in technology and their innate ability to overcome contact angle hysteresis \cite{Lei2016}, similar droplets of two (or more) liquid components of different volatilities are therefore prime candidates for applications that require control over droplet motion on surfaces. When it comes to directional droplet manipulation, however, existing techniques rely on large gradients of surface energy \cite{Chaudhury1992,Daniel2001}, on carefully engineered substrates \cite{Pollack2002,Jiang2016,Tian2016}, or on tailored trapping potentials \cite{Cira2015}, thus limiting the level of control that can be achieved over droplet motion and its applicability on pristine substrates, for example, for printing technology. 

Here, we demonstrate the contactless long-range 2D manipulation of binary droplets of water and propylene glycol on solid pristine surfaces. These droplets spontaneously move in response to small imbalances of surface tension gradients induced by the presence of an external localized source of water vapor. We show with an analytical model that the dissipative nature of the driving forces at play ($<{\rm \mu N}$) is paramount to capture the observed dual response (from attractive to repulsive) of the droplet with distance from the source. Our robust understanding of the underlying motion mechanism allows us to showcase its adaptability in a range of potential applications in materials science, including pattern formation, the printing of chemical gradients and the mixing of reactive materials.

\begin{figure}
\centering
\includegraphics[width=0.66\textwidth]{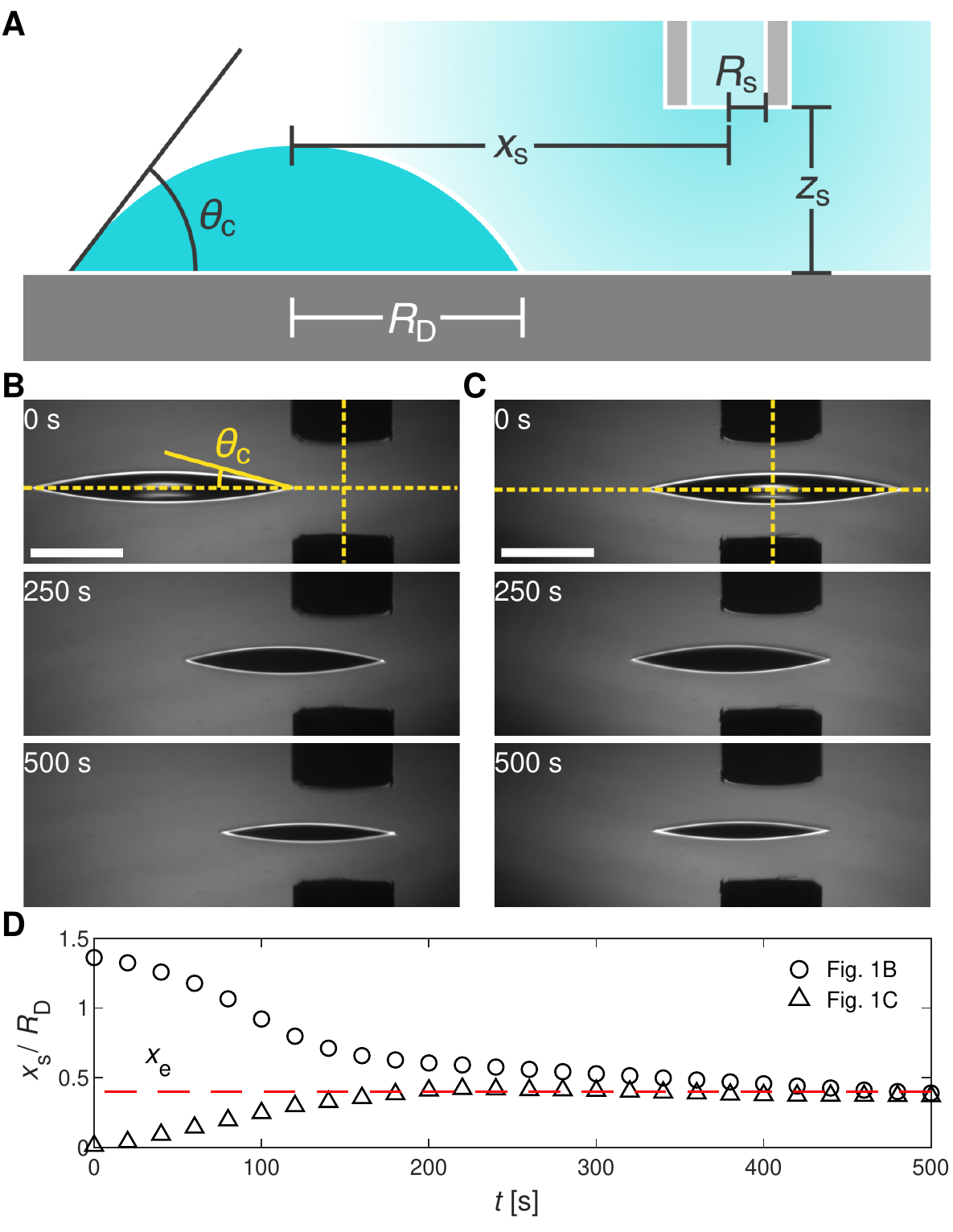}
\caption{{\bf Binary droplets under an external vapor source: attraction vs. repulsion.} ({\bf A}) Schematics of the contactless manipulation of binary droplets on solid substrates with an external localized vapor source: a binary droplet (radius $R_{\rm D}$; contact angle $\theta_{\rm c}$) is placed at a distance $x_{\rm s}$ from a blunt needle (inner radius $R_{\rm s}$) from where water vapor diffuses; $z_{\rm s}$ is the distance between source and substrate. All experiments are performed within an environmental chamber with controlled temperature ($T = 21 \pm 0.5 \, \degree {\rm C}$) and relative humidity ($RH = 50 \pm 5\%$). ({\bf B-C}) Time sequences showing a $0.5 \, {\rm \mu L}$  binary droplet of water and propylene glycol ($R_{\rm D} = 1.38 \pm 0.06 \, {\rm mm}$; $\theta_{\rm c} = 12.5 \pm 0.7\degree$; mole fraction of water $x_{\rm H_2O} = 0.95$) being ({\bf B}) attracted to or ({\bf C}) repelled from the source ($R_{\rm s} = 350 \, {\rm \mu m}$) depending on the initial separation $x_{\rm s}$:  ({\bf B}) $x_{\rm s} = 2 \, {\rm mm}$ and ({\bf C}) $x_{\rm s} = 0$. The horizontal and vertical dashed lines highlight the substrate and the center of the vapor source, respectively. Scale bar: $1 \, {\rm mm}$. ({\bf D}) Time evolution of $x_{\rm s}$ converging to the same radial distance $x_{\rm e}$ (dashed line) from the source for the droplets in {\bf B} (circles) and {\bf C} (triangles).} 
\label{fig:fig1}
\end{figure}

\subsection*{Motion of binary droplets under an external vapor source}

In the most basic configuration, a $0.5 \, {\rm \mu L}$ binary droplet of water and propylene glycol (mole fraction of water $x_{\rm H_2O} = 0.95$) is deposited on a clean glass slide at a distance $x_{\rm s}$ from an external localized source of water vapor (Fig. \ref{fig:fig1}A) \cite{Supplementary}. As evaporation is faster at the droplet's edges \cite{Deegan1997} and propylene glycol (PG) is less volatile and with a lower surface tension than water, a radially symmetric gradient of surface tension $\gamma$ forms on the droplet's free surface \cite{Cira2015,Karpitschka2017}. The resulting inward Marangoni stresses prevent spreading \cite{Pesach1987}, so that, due to its ongoing evaporation, the droplet features a contact angle ($\theta_{\rm c} = 12.5 \pm 0.7\degree$) higher than either of the pure liquids and a reduced hysteresis \cite{Karpitschka2017}. When $x_{\rm s} \to \infty$ (i.e. in the absence of the source), the two-component droplet is stationary on a flat surface; however, in the presence of the source, because of the reduced hysteresis, the binary droplet can instead move if the contact line is not pinned as in our case. In particular, when placed afar ($x_{\rm s} = 2 \, {\rm mm}$, Fig. \ref{fig:fig1}B, the droplet experiences an attractive force towards the source. This attractive behavior is consistent with a relative larger local increase in humidity at the droplet's edge nearest to the source, which generates motion by slowing down evaporation and comparatively increasing the value of surface tension at that edge with respect to the opposite side \cite{Cira2015,Man2017}. Based on these considerations on surface tension, one could naturally expect a stable equilibrium position to appear directly beneath the source's center ($x_{\rm s} = 0$), as any displacement from $x_{\rm s} = 0$ would induce a restoring force. Surprisingly, $x_{\rm s} = 0$ is an unstable equilibrium position from which the droplet gets easily repelled to an off-centered radial distance $x_{\rm e}$ (Fig. \ref{fig:fig1}C, where the droplet consistently settles near the end of the evaporation whether it starts from afar the vapor source or below it (Fig. \ref{fig:fig1}D).

\subsection*{Dynamics of droplet's motion}

\begin{figure}
\centering
\includegraphics[width=0.65\textwidth]{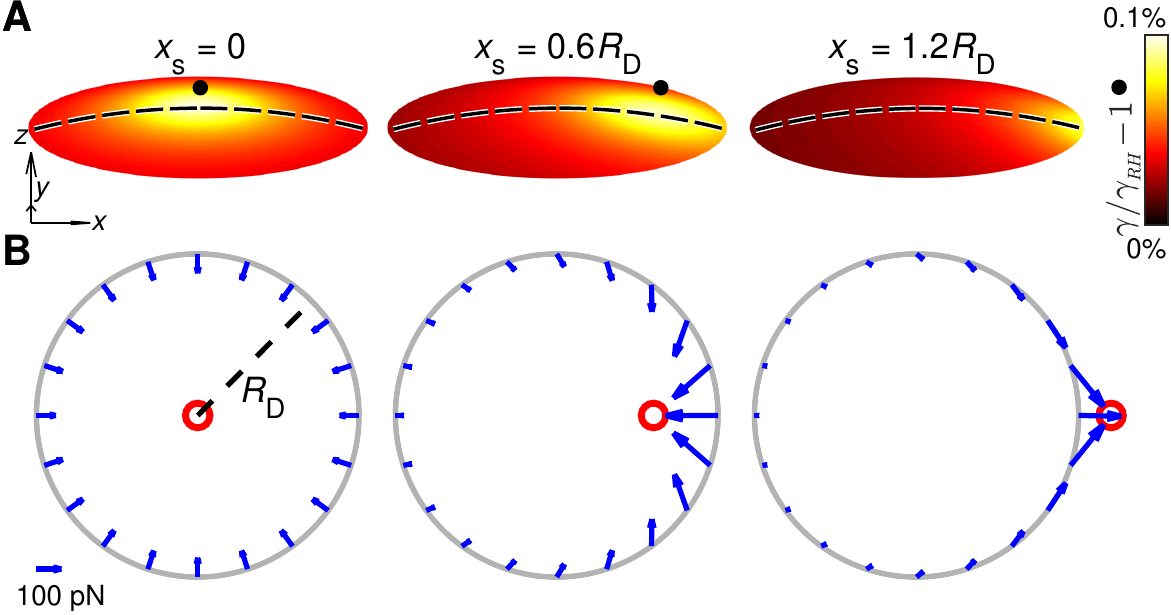}
\caption{{\bf Forces induced on binary droplets under an external vapor source.}  An external vapor source induces gradients of surface tension $\gamma$ on a binary droplet's free surface. The corresponding Marangoni flows generate forces on the droplet's contact line, ultimately leading to its motion towards or away the vapor source. ({\bf A}) Calculated change of surface tension $\gamma$ along the droplet's free surface due to vapor sources (black dots, $R_{\rm s} = 350 \, {\rm \mu m}$) placed directly above ($x_{\rm s} = 0$), with a slight offset ($x_{\rm s} = 0.6 R_{\rm D}$) and afar from ($x_{\rm s} = 1.2 R_{\rm D}$) the droplet when evaporation starts. The change in $\gamma$ is given relative to a reference value $\gamma_{ RH}$ calculated removing the effect of the source \cite{Supplementary}. Dashed lines: meridians through the droplets' apices. The coordinate unit vectors correspond to $0.5 \, {\rm mm}$. ({\bf B}) Calculated local force (blue arrows) along the contact line (gray circles) due to vapor sources as in {\bf A} (red circles). Force vectors, shown every $\frac{\pi}{10}$, are integrated over $\frac{\pi}{500}$ intervals along the contact line.}
\label{fig:fig2}
\end{figure}

To efficiently harness this mechanism, we have developed a simplified analytical model to  better understand our experimental observations. In fact, this duality (attraction vs. repulsion) in the interaction between droplet and source cannot be simply explained by arguments purely based on differences of surface tension on opposing sides of the droplet, for which the droplet would always move towards the vapor source \cite{Man2017}. Instead, as can be seen in Fig. \ref{fig:fig2}, both the magnitude and directionality of the net force experienced by the droplet can be modulated when we account for the gradient of surface tension along its contact line. 

To a first approximation, the whole process can be understood assuming a steady state for the diffusion of water vapor from the source towards the droplet's free surface (Supplementary Text). In the presence of the vapor source, aqueous vapor saturates the atmosphere within a blunt needle (vapor pressure $p_{\rm s} = 2.49 \, {\rm kPa}$ at $T = 21 \, \degree {\rm C}$) and diffuses towards the droplet's free surface, where it induces a local decrease in the evaporation rate proportional to the local water partial pressure, $p_{\rm H_2O}(r,\theta,h)$, with $r$, $\theta$ and $h(r)$ being variables describing the free surface in cylindrical coordinates (Fig. S\ref{fig:figS1}A and Supplementary Text). We assume that this local decrease in evaporation rate induces local variations in composition $\chi_{\rm H_2O}(r,\theta,h)$, and hence in surface tension $\gamma(r,\theta,h)$, along the free surface with respect to the bulk composition of the evaporating droplet $x^{\rm b}_{\rm H_2O}$ (Supplementary Text). As a consequence, gradients of surface tension form over the droplet's free surface that drive the formation of Marangoni flows towards the areas of higher $\gamma$ \cite{Hu2005}, i.e. where the evaporation is slower. The calculations in Fig. \ref{fig:fig2}A show how the position of this maximum shifts along the droplet's free surface following the position of the source, moving from the droplet's apex for $x_{\rm s} = 0$ to its contact line when the source is far away ($x_{\rm s} = 1.2 R_{\rm D}$). For no displacement ($x_{\rm s} = 0$), the gradient in surface tension, and hence the corresponding Marangoni flows along the free surface, are radially symmetric and pointing inward towards the droplet's apex (Fig. S\ref{fig:figS2}). When the source is displaced towards one edge, the flows in the droplet become radially asymmetric \cite{Malinowski2018}, strengthening under the source at first due to a steeper gradient in surface tension (Fig. S\ref{fig:figS2}). When the source is far away, the flows coming from the distal edge predominate instead due to the confined geometry of the droplet. More importantly, between these two cases, the gradient in surface tension at the edge closer to the source changes directionality (Fig. S\ref{fig:figS2}). This change intuitively justifies the dual response of the droplet's motion to the position of the source (Fig. \ref{fig:fig1}).

To formalize this intuition, we can integrate the viscous stress induced by the flows within the droplet on the liquid-solid interface to obtain a dissipative driving force in the direction of motion (Supplementary Text):
\begin{equation}\label{eq1}
 	F^{\gamma}_x =  \int_{0}^{2\pi} \frac{R_{\rm D}}{2} \big( R_{\rm D} \frac{\partial \gamma}{\partial r} \cos \theta - \frac{\partial \gamma}{\partial \theta} \sin \theta \big) \, d\theta.
\end{equation}
This force primarily acts on the contact line, is a function of the gradient of surface tension, and has typical values $< {\rm \mu N}$ consistent with previous reports (Fig. S\ref{fig:figS3}) \cite{Cira2015}. Figure \ref{fig:fig2}B shows how the integrated function (with units of force) varies along the contact line for three different source positions at the start of the evaporation. This quantity always points towards the source and intensifies in the parts of the contact line closer to it. These two considerations alone can formally explain our main observations in Fig. 1. When the source is just above the droplet's apex ($x_{\rm s} = 0$), there is no net force because of symmetry; this is, however, an unstable configuration as any small displacement generates repulsion ($x_{\rm s} = 0.6 R_{\rm D}$). When the source crosses the contact line ($x_{\rm s} = 1.2 R_{\rm D}$), the droplet starts experiencing a net attractive force instead, and a stable equilibrium point is formed at the crossover between attraction to and repulsion from the source. 

\begin{figure}
\centering
\includegraphics[width=0.65\textwidth]{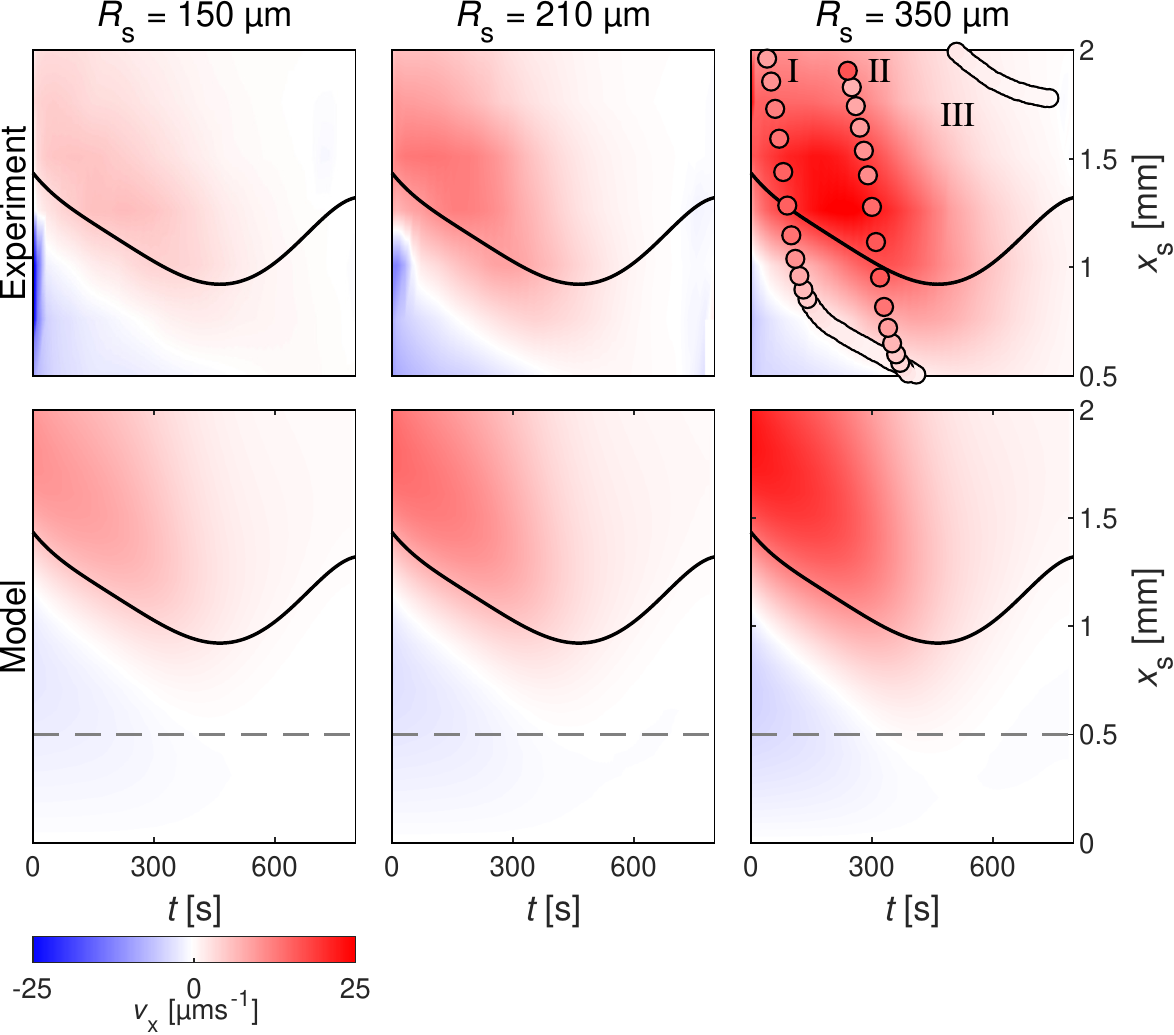}
\caption{{\bf Droplet's velocity in experiments and model.} Droplet's mean velocity $v_{x}$ with distance $x_{\rm s}$ from the source and time $t$ from when evaporation starts for increasing $R_{\rm s}$ ($R_{\rm s} = 150 \, {\rm \mu m}$, $R_{\rm s} = 210 \, {\rm \mu m}$ and $R_{\rm s} = 350 \, {\rm \mu m}$) in experiments and simulations. Mean values are averages of at least 5 different experiments with a standard deviation of $0.5 \, {\rm \mu ms^{-1}}$ at $R_{\rm s} = 150 \, {\rm \mu m}$, $0.8 \, {\rm \mu ms^{-1}}$ at $R_{\rm s} = 210 \, {\rm \mu m}$ and $1.5 \, {\rm \mu ms^{-1}}$ at $R_{\rm s} = 350 \, {\rm \mu m}$. These experiments were performed by continuously moving the stage to keep the distance between source and droplet fix at an initial set value $x_{\rm s}$ \cite{Supplementary}. The solid lines represent the time evolution of the droplet's radius $R_{\rm D}$ (Fig. S\ref{fig:figS1}B). In the simulations, the dashed lines delimit the part where corresponding experiments are available. In the experiments for $R_{\rm s} = 350 \, {\rm \mu m}$, the circles represent three trajectories of individual droplets (circles) moving towards the source from $x_{\rm s} = 2 \, {\rm mm}$ (as in Fig. \ref{fig:fig1}B) at different times (Fig. S\ref{fig:figS4}): (I) $30 \, {\rm s}$, (II) $250 \, {\rm s}$ and (III) $500 \, {\rm s}$ from the beginning of their evaporation. The color code of each circle represents the droplet's velocity at a given time.}
\label{fig:fig3}
\end{figure}

Finally, by equating Eq. \ref{eq1} to the viscous drag force on the moving droplet (Supplementary Text), the velocity $v_x$ of the droplet in the vapor field generated by the source can be calculated as
\begin{equation}\label{eqS8}
 	v_x = \frac{\theta_{\rm c}}{12\pi\eta \ell_{\rm n}} \int_{0}^{2\pi} \big( R_{\rm D} \frac{\partial \gamma}{\partial r} \cos \theta - \frac{\partial \gamma}{\partial \theta} \sin \theta \big) \, d\theta,
\end{equation}
where $\eta$ is the dynamic viscosity of the mixture at a given composition and $\ell_{\rm n} = 11.2$ is a cutoff constant (Supplementary Text). Typical experimental values of $v_x$ range between a few and a few tens of ${\rm \mu m s^{-1}}$, and are well reproduced by our model (Fig. \ref{fig:fig3}).

As the droplet's geometry and composition are changing due to evaporation (Fig. S\ref{fig:figS1} and Supplementary Text), both force and velocity vary in time too (Figs. \ref{fig:fig3} and S\ref{fig:figS3}). The experiments in Fig. \ref{fig:fig3} show how the droplet's velocity evolves in time and with the distance $x_{\rm s}$ from the source as a function of the source size $R_{\rm s}$. Qualitatively, these maps show the same common features, which are well reproduced by our model. At the beginning of the evaporation process, a small region of null speed and force (white, Figs. \ref{fig:fig3} and S\ref{fig:figS3}) in the proximity of the droplet's edge separates a repulsive area (blue) from an attractive one (red). As the droplet's radius is initially shrinking (Fig. S\ref{fig:figS1}B), the stable equilibrium point of null speed, and hence the limit of the repulsive zone, shift towards the droplet's center with time, giving way to the attractive zone. When moving away from the droplet's edge, this attractive region presents a time-dependent maximum around $\langle x_{\rm s} \rangle \approx 1.3 R_{\rm D}$ as the strength of the attractive forces fades when the distance from the source increases due to weaker vapor fields around the droplet. Although the force strengthens towards the end of the evaporation (Fig. S\ref{fig:figS3}), speed drops eventually to zero as the droplet becomes richer in PG over time  (and its viscosity increases), so that significantly higher forces are needed to set it in motion. As can be seen in the simulated maps (Figs. \ref{fig:fig3} and S\ref{fig:figS3}), a region of null speed and force emerges also for $x_{\rm s} \approx 0$. As previously observed (Fig. \ref{fig:fig1}), this represents an unstable equilibrium point, which is indeed surrounded by a region of repulsive forces (Fig. S\ref{fig:figS3}). Qualitative differences (also reproduced by our model) can be attributed to the source size, i.e. the strength of its influence at a given $x_{\rm s}$ (Fig. \ref{fig:fig3}). In particular, larger sources exert larger attractive forces (Fig. S\ref{fig:figS3}) and, as a consequence, droplets move faster and experience the source influence from further away and for longer during their evaporation. It is important to note that, to keep a fixed distance $x_{\rm s}$ between droplet and source over time, these experiments were performed by continuously moving the microscope stage to compensate for the droplet's motion \cite{Supplementary}. Interestingly, when droplets are free to move on the substrate as in Fig. \ref{fig:fig1}, they follow trajectories on the maps in Fig. \ref{fig:fig3} with a time-varying velocity consistent with the underlying functional form of these velocity surfaces (Fig. S\ref{fig:figS4}) until they settle in the stable region of null speed.

\begin{figure}[h!]
\centering
\includegraphics[width=0.6\textwidth]{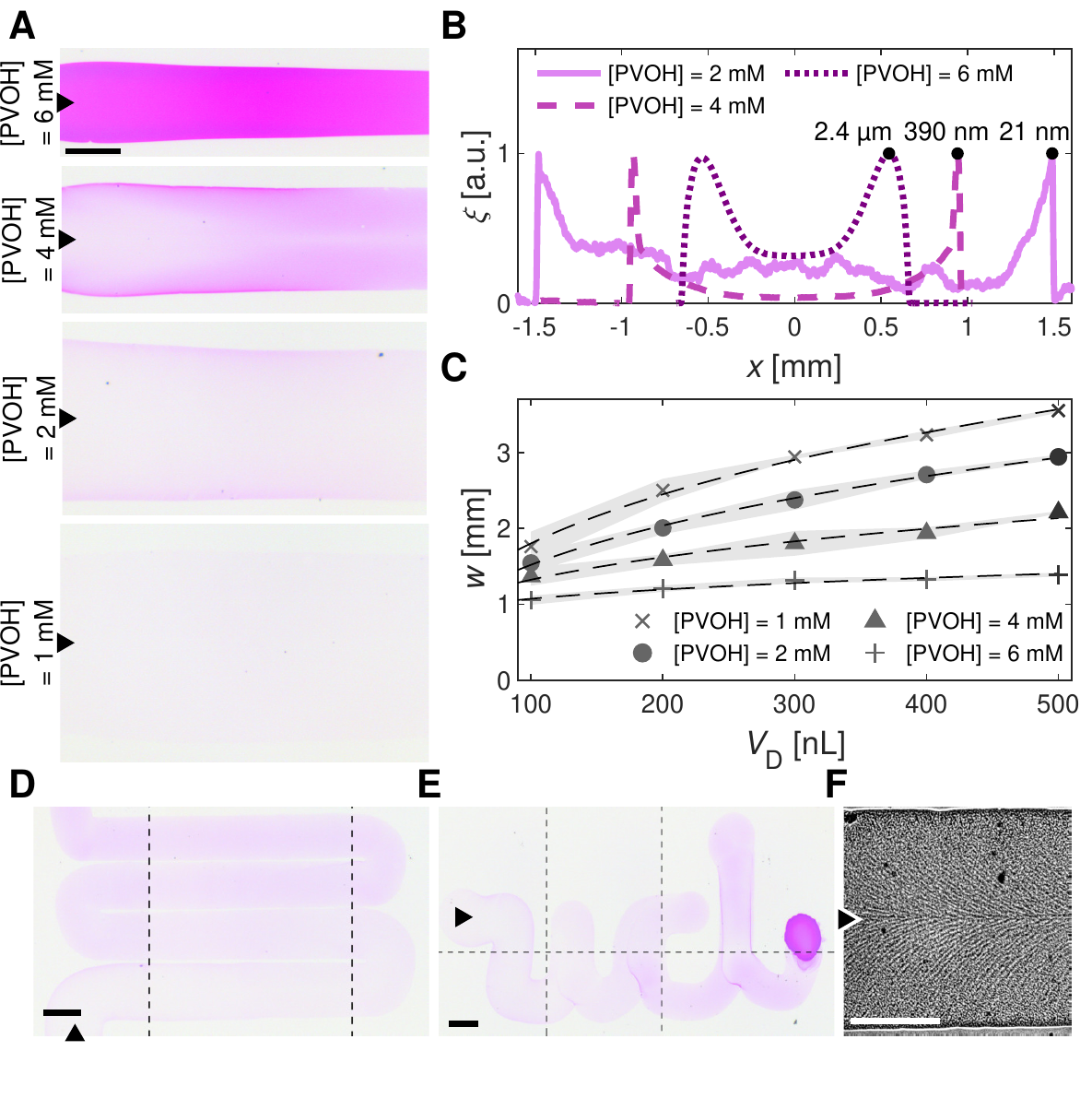}
\caption{{\bf Printing with moving droplets.} ({\bf A}) Photographs of linear polymer deposits (PVOH) from moving water/PG droplets ($V_{\rm D} = 0.5 \, {\rm \mu L}$, $x_{\rm H_2O}$ = 0.95) with decreasing PVOH concentrations. ({\bf B}) Typical height profiles for the deposits in ({\bf A}) \cite{Supplementary}. ({\bf C}) Deposit maximum line width $w$ as a function of droplet's volume and PVOH concentration, averaged across at least three deposits. The dashed lines are fits to the experimental data showing the proportionality between $w$ and $V_{\rm D}^{1/3}$. ({\bf D-E}) Stitched photographs of PVOH deposits ($[{\rm PVOH}] = 2 \, {\rm mM}$) from moving water/PG droplets ($V_{\rm D} = 0.1 \, {\rm \mu L}$, $x_{\rm H_2O}$ = 0.95) guided along 2D patterns to form ({\bf D}) a serpentine with an $\approx 30 \, {\rm \mu m}$ interline spacing and ({\bf E}) the letters ``ucl''. ({\bf F}) Example photograph (enhanced with an edge-aware filter for contrast) of alignment in linear polymer deposits ($[{\rm PEG}] = 100 \, {\rm mM}$) from moving water/PG droplets ($V_{\rm D} = 0.5 \, {\rm \mu L}$, $x_{\rm H_2O}$ = 0.95) guided at $8 \, {\rm \mu ms^{-1}}$ \cite{Supplementary}. Higher speeds are shown in Fig. S\ref{fig:figS6}. All PVOH droplets contain rhodamine B for visualization. The background was subtracted from all colour images. In the photographs, black triangles indicate the direction of motion. Scale bars: 1 mm.}
\label{fig:fig4}
\end{figure}

\subsection*{Applications of moving droplets under an external vapor source}

\begin{figure}[h!]
\centering
\includegraphics[width=0.66\textwidth]{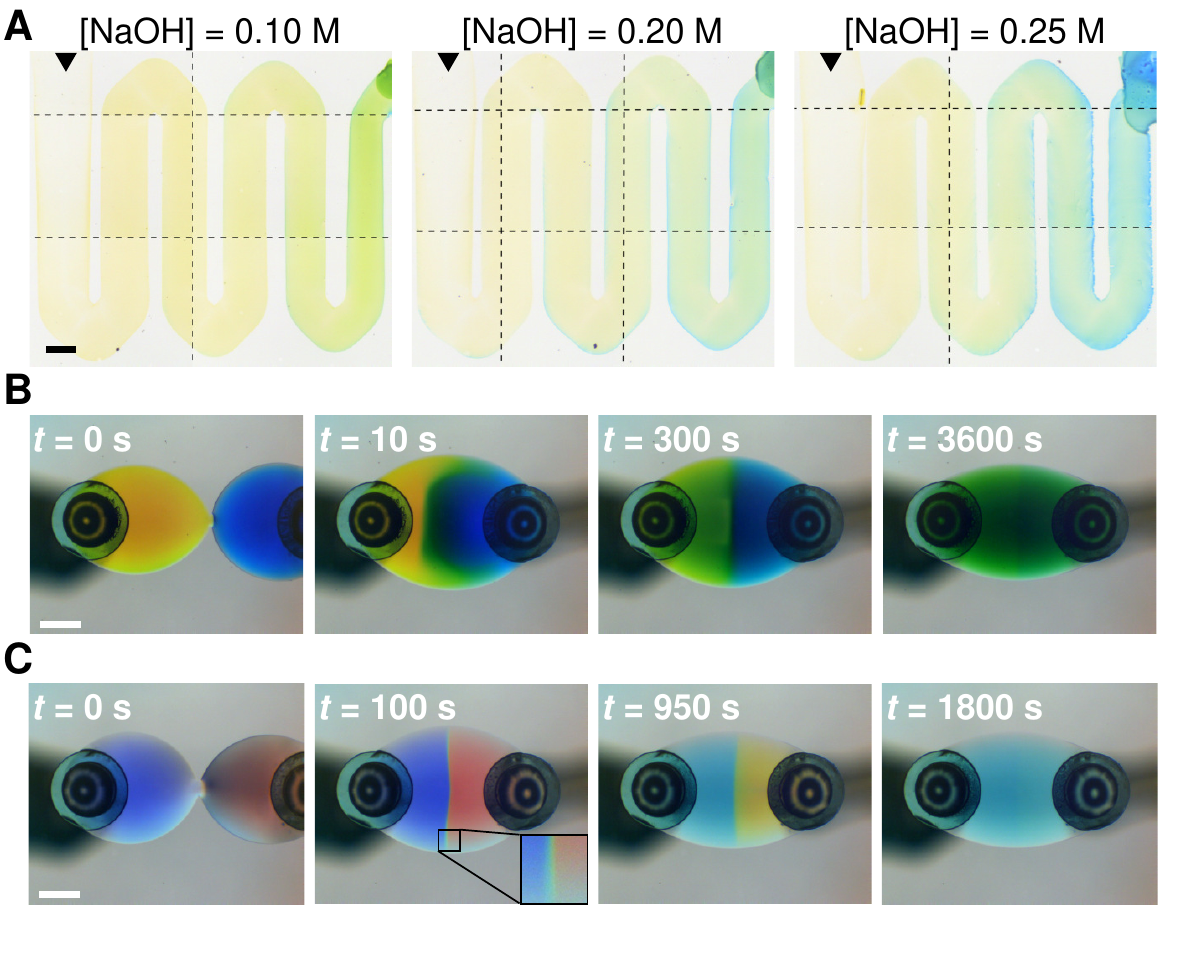}
\caption{{\bf Chemical reactors with moving droplets.} ({\bf A}) Stitched photographs of color gradients in polymer deposits obtained by retracing water/PG droplets ($V_{\rm D}$ = 0.25 $\mu$L, $x_{\rm H_2O}$ = 0.95, [PVOH] = 4 mM) containing 30 mM bromothymol blue over previous deposits by droplets ($V_{\rm D}$ = 0.15 $\mu$L, $x_{\rm H_2O}$ = 0.95, [PVOH] = 2 mM) containing different NaOH concentrations \cite{Supplementary}. Black triangles indicate the direction of motion. ({\bf B-C}) Time sequences showing coalescence of two water/PG droplets ($V_{\rm D} = 0.5 \, {\rm \mu L}$, $x_{\rm H_2O}$ = 0.85) controlled by two vapor sources ($R_{\rm s} = 640 \, {\rm \mu m}$) \cite{Supplementary}. Droplets start coalescing at $0 \, {\rm s}$ and remain visibly compartmentalized for a long time as confirmed by the flow lines in Fig. S\ref{fig:figS7}. In {\bf B}, the droplets contain 100 mM NaOH and a dye, either methyl red (yellow, left) or bromothymol blue (blue, right); their contents mix as evidenced by the coalesced droplet turning green. In {\bf C}, the droplets contain a pH universal indicator and either 100 mM NaOH (left) or 100 mM HCl (right); upon coalescing, an acid-base neutralization reaction occurs (inset) until the coalesced droplet turns uniform as the indicator shows qualitatively. The background was subtracted from all images. Scale bars: 1 mm.}
\label{fig:fig5}
\end{figure}

Next, we demonstrate the performance and versatility of our method for possible applications, ranging from printing and depositing materials to controlling reactions in space and time (Figs. \ref{fig:fig4} and \ref{fig:fig5}). Figure \ref{fig:fig4} shows deposits from water/PG droplets containing different polymer concentrations. The droplets were guided with a large vapor source ($R_{\rm s} = 640 \, {\rm \mu m}$) placed at their leading edge to overcome the higher viscosity induced by the presence of the polymer \cite{Supplementary}. Figure \ref{fig:fig4}A shows polymer trails left behind by moving droplets containing different concentrations of polyvinyl alcohol (PVOH), a polymer used for coating and printing applications \cite{Goodship2009}. Visually, higher concentrations of polymer lead to thicker deposits with narrower line widths as a result of an increase in the droplet's overall viscosity and contact angle $\theta_{\rm c}$. The height profiles in Fig. \ref{fig:fig4}B and the measurements in Fig. \ref{fig:fig4}C confirm these observations quantitatively. Figure \ref{fig:fig4}B also shows that typical measured thicknesses span at least two orders of magnitude with deposits as thin as $\approx 5 \, {\rm nm}$. For a given PVOH concentration, the resolution achievable by this printing technique in terms of average line width depends on the characteristic length of the droplet as $w \propto V_{\rm D}^{\frac{1}{3}}$ \cite{doi2013}, with smaller droplets thus depositing thinner lines  (Fig. \ref{fig:fig4}C and Fig. S\ref{fig:figS5}). The addition of a second degree of freedom to the in-plane displacement of the droplet allows for printing arbitrary patterns in two-dimensions such as sinuous lines (Fig. \ref{fig:fig4}D) and the cursive letters ``ucl'', notably including double-pass portions (Fig. \ref{fig:fig4}E). Finally, Figs. \ref{fig:fig4}F and S\ref{fig:figS6} demonstrate the possibility of controlling the orientation of a polycrystalline polymer, such as PEG \cite{Tien2014}, thus leading to the formation of patterns within the deposit itself with $\approx 100 \, {\rm nm}$ high features (Fig. S\ref{fig:figS6}B). In fact, PEG quickly supersaturates in the deposit printed by the droplet (due to the evaporation of its more volatile components) and starts to crystallise as soon as a defect is generated \cite{Granasy2005}. The resulting polycrystalline phase forms physical ridges perpendicular to a moving front whose shape depends on the droplet's speed and determines the topography of the final pattern (Fig. S\ref{fig:figS6}): for slower moving droplets, the front matches the droplet's circular shape leaving a scallop shell pattern behind, while, as the droplet speed increases, the moving front deforms into a triangular shape which leaves a herringbone structure behind instead.

Beyond depositing material in a controlled manner (Fig. \ref{fig:fig4}), our technique also allows redissolving previously deposited material within a second droplet, thus facilitating the controllable deposition of chemical reactions in space and time. For example, Fig. \ref{fig:fig5}A shows the final deposits left behind by water/PG droplets containing a pH indicator (bromothymol blue) after they retrace previous deposits by basic water/PG droplets with varying concentrations of NaOH. As a pH indicating droplet (initially yellow) retraces a previous deposit at constant speed (here $v_x = 25 \, {\rm \mu ms^{-1}}$), it uptakes NaOH from the substrate, its pH increases, and its color turns from yellow to blue through green, thus printing a color gradient. The steepness of this gradient, and thus the spatial variation of the printed colors, can be regulated through the rate of uptake of the first deposited reactant (here NaOH), e.g., by varying its concentration in the first droplet (Fig. \ref{fig:fig5}A).  

Finally, Figs. \ref{fig:fig5}B and C show the simultaneous control of different droplets with more than one vapor source, which can be useful to trigger the mixing and the reaction of materials dissolved or suspended within them in space and time. As can be seen in Fig. \ref{fig:fig5}B, the coalescence of two droplets can be guided by controlling the separation distance between two vapor sources \cite{Supplementary}. The flows in the resulting droplet are significantly different from the radially symmetric flows in a standard sessile binary droplet \cite{Cira2015}, and, after flowing outwards along the liquid-liquid interface between the two coalesced droplets, recirculate towards the closest source in each compartmentalized quarter (Fig. S\ref{fig:figS7}). As a consequence, after solute exchange has occurred at the interface predominantly via diffusion, mixing is promoted within each quarter by these flows until the coalesced droplet becomes essentially uniform in composition. Beyond mixing, chemical reactions can also be implemented by the same principle. As an example, Fig. \ref{fig:fig5}C shows the acid-based neutralization reaction of NaOH and HCl: before coalescence, the two droplets are respectively very acidic (${\rm pH \approx 1}$) and basic  (${\rm pH \approx 13}$); on coalescence, neutralization starts to occur at the interface where a gradient of pH can be observed as highlighted by the full color spectrum of the pH indicator; as the reaction proceeds, the coalesced droplet eventually reaches neutral pH uniformly. 

\subsection*{Conclusions}

These examples illustrate the robustness and versatility of our novel method for the contactless, long-range 2D manipulation of droplets on solid substrates. The local and tunable nature of the perturbation introduced by the vapor source makes it ideal for fundamental studies of droplet motion \cite{Man2017,Benusiglio2018,Sadafi2019}. Due to its high sensitivity, our technique's principle could be adapted to develop force (from ${\rm nN}$ to ${\rm \mu N}$) and gas sensors. Similarly, beyond the sample applications demonstrated here, we envisage that our technique could be employed for coating applications, in printable electronics and optoelectronics, for manufacturing \cite{Li2017,Li2018} as well as for the development of chemical reactors, diagnostic tools and bioassays based on small volume liquid handling \cite{Zhang2013,Han2018,Yang2018,Xiao2018,Zhang2018}.


\bibliographystyle{Science}

\section*{Acknowledgments}

We are grateful to \'Alvaro Marin, Jacco Snoeijer and Giovanni Volpe for fruitful discussions about the model. We acknowledge the COST Action MP1305 ``Flowing Matter'' for providing several meeting occasions. {\bf Funding:} Giorgio Volpe (GV) acknowledges funding from the HEFCE's Higher Education Innovation Fund (KEI2017-05-07). Robert Malinowski (RM) and Ivan P. Parkin (IPP) acknowledge funding from EPSRC (EP/G036675/1). {\bf Authors Contributions:} GV conceived the idea for the work. GV and RM designed the experiments and developed the model. RM performed the experiments and the numerical simulations. GV and RM analysed the data and wrote the manuscript. GV and IPP supervised the project. All authors discussed the results and revised the final version of the manuscript. {\bf Competing interests:} The authors have no competing interests. {\bf Data and materials availability:} All data in support of this work is available in the manuscript or the supplementary materials. Further data, codes or materials are available from the corresponding author upon reasonable request. 

\section*{Supplementary materials}

Materials and Methods\\
Supplementary Text\\
Figs. S1 -- S7\\
References \textit{(44-53)} \\

\clearpage

\section*{Supplementary materials}

\subsection*{Materials and Methods}

\subsubsection*{Materials}

Glass slides (AAAA000001\#\#02E, Menzel Gläser) and glass capillaries (G119/02, Samco) were purchased from VWR. The largest needle used here ($R_{\rm s}$ = 350 $\mu$m) was purchased from Fisher Scientific (BD 301750). The smaller blunt needles ($R_{\rm s}$ = 150 $\mu$m, KDS2312P; $R_{\rm s}$ = 210 $\mu$m, KDS2512P) were purchased from Farnell. Needles were removed from their casing by soaking them in acetone ($\geq$ 99.8\%, Sigma-Aldrich). Propylene glycol ($\geq$ 99.5\%, Sigma-Aldrich), ethanol ($\geq$ 99.8\%, Fisher Scientific), aqueous hydrochloric acid (37\%, VWR), sodium hydroxide (analytical grade, Fisher Scientific),  poly(vinyl alcohol) ($M_{\rm W}$ = 9,000-10,000 gmol\textsuperscript{-1}, 80\% hydrolysed, Sigma-Aldrich),  polyethylene glycol  4000 (BDH Chemicals), rhodamine B (Acros Organics), thymol blue (Fisons),  methyl red (Alfa Aesar), bromothymol blue (BDH Chemicals), and phenolphthalein (Sigma-Aldrich) were used as received without further purification. A suspension of monodisperse polystyrene particles ($5 \, {\rm \mu m}$ diameter, 10 wt\%) was purchased from microParticles GmbH. DI water (resistivity $> 18 \, {\rm M \Omega .cm}$) was obtained from a Milli-Q water purification system. 

Water/propylene glycol (PG) stock solutions were prepared by combining the two components in the correct mass to give the desired mole fraction ($x_{\rm H_2O}$). For example, $x_{\rm H_2O}$ = 0.95 was made by mixing propylene glycol (2.0 g, 26.3 mmol) with DI water (9.0 g, 500 mmol). These solutions were used as a base for the preparation of the following stock solutions from which droplets were made. A 10 mM poly(vinyl alcohol) stock solution was prepared by dissolving  poly(vinyl alcohol) (95 mg) in water/PG (1 mL, $x_{\rm H_2O}$ = 0.95). A 20 mM rhodamine B stock solution was made by dissolving rhodamine B (48 mg) in water/PG (5 mL, $x_{\rm H_2O}$ = 0.95). A 100 mM polyethylene glycol stock solution was prepared by dissolving  polyethylene glycol 4000 (0.20 g) in water/PG (0.5 mL, $x_{\rm H_2O}$ = 0.95). A 2 M NaOH stock solution was prepared by dissolving NaOH (0.20 g) in water/PG (2.5 mL, $x_{\rm H_2O}$ = 0.95). A 1 M NaOH stock solution was prepared by dissolving NaOH (0.20 g) in water/PG (5 mL, $x_{\rm H_2O}$ = 0.85 or $x_{\rm H_2O}$ = 0.95). A 1 M HCl stock solution was made by combining aqueous HCl (37\%, 0.492 g) with PG (0.231 g) and then diluting to 5 mL with water/PG ($x_{\rm H_2O}$ = 0.85). A neutral 60 mM bromothymol blue stock solution was made by dissolving bromothymol blue (37.5 mg) in water/PG (1 mL, $x_{\rm H_2O}$ = 0.95) and then by neutralizing to dark yellow with a few drops of the 1M NaOH stock solution ($x_{\rm H_2O}$ = 0.95). A basic 5 mM bromothymol blue dye stock solution was made by dissolving bromothymol blue (6.24 mg) in a 100 mM NaOH solution (2 mL, $x_{\rm H_2O}$ = 0.85) obtained from the 1M stock solution ($x_{\rm H_2O}$ = 0.85). A 10 mM methyl red dye solution was made by dissolving methyl red (2.69 mg) in a 100 mM NaOH solution (1 mL, $x_{\rm H_2O}$ = 0.85) obtained from the 1M stock solution ($x_{\rm H_2O}$ = 0.85). A pH indicating stock solution, similar to Yamada's universal indicator\cite{Foster1937}, was prepared by dissolving thymol blue (1.25 mg), methyl red (3.13 mg), bromothymol blue (15.0 mg) and phenolphthalein (25.0 mg) in propylene glycol (8.55 g). A few drops of 1 M NaOH stock solution ($x_{\rm H_2O}$ = 0.85) were added to help improve the solubility of the dyes and make the solution pH neutral (green). DI water (11.45 g) was then added, and the solution was diluted to 25 mL using a water/PG mixture ($x_{\rm H_2O}$ = 0.85).

\subsubsection*{Sample Preparation}

Glass slides were cleaned by sonicating sequentially in a $2 \, {\rm M}$ NaOH ethanolic solution for 10 minutes, DI water for 5 minutes, a $1 \, {\rm M}$ HCl aqueous solution for 10 minutes and, finally, DI water for 5 minutes three times. Each slide was then dried by withdrawing it from the last water bath while exposing its surface to ethanol vapor (Marangoni drying) \cite{Leenaars1990}. Any remaining water was removed with a nitrogen gun. In Figs. \ref{fig:fig1}, \ref{fig:fig3}, S\ref{fig:figS1} and S\ref{fig:figS4}, binary droplets were prepared from the water/PG stock solution ($x_{\rm H_2O}$ = 0.95) without additives. In Figs. \ref{fig:fig4}A to E and S\ref{fig:figS5}, droplets were generated from solutions of varying poly(vinyl alcohol) concentration ($1 \, {\rm mM} \le [{\rm PVOH]} \le 5 \, {\rm mM}$) dyed with 10 mM rhodamine B. These solutions were prepared by diluting the 10 mM stock solution of poly(vinyl alcohol) with water/PG ($x_{\rm H_2O}$ = 0.95) to $2[{\rm PVOH]}$ and then mixing with the 20 mM rhodamine B stock solution in equal parts. In Fig. \ref{fig:fig4}A and B, droplets with 6 mM poly(vinyl alcohol) and 10 mM rhodamine B were instead prepared by directly dissolving poly(vinyl alcohol) (28.5 mg) in 20 mM rhodamine B stock solution (2.5 mL) and water/PG (2.5 mL, $x_{\rm H_2O}$ = 0.95). In Figs. \ref{fig:fig4}F and S\ref{fig:figS6}, droplets containing PEG were directly prepared from the 100 mM polyethylene glycol stock solution. In Fig. \ref{fig:fig5}A, NaOH depositing droplets ([PVOH] = 2 mM) were prepared by diluting the 2 M stock NaOH solution to $2[{\rm NaOH]}$ with water/PG ($x_{\rm H_2O}$ = 0.95) and then mixing with a solution of  4 mM $[{\rm PVOH]}$ in equal parts. In Fig. \ref{fig:fig5}B, dye containing droplets were prepared from the stock solutions of 5 mM bromothymol blue and 10 mM methyl red. The factor of 2 between the concentrations of the dyes is to compensate for the difference in their extinction coefficients. In Fig. \ref{fig:fig5}C, pH indicating droplets containing 100 mM NaOH (or 100 mM HCl) were prepared by diluting the 1 M NaOH (or 1 M HCl) stock solution ($x_{\rm H_2O}$ = 0.85) to 200 mM with water/PG ($x_{\rm H_2O}$ = 0.85), and then combining in equal parts with the pH indicating stock solution. In Fig. S\ref{fig:figS7}, droplets containing microparticles for flow visualization were prepared by diluting the stock aqueous suspension of 5 $\mu$m polystyrene particles (5$\mu$L, 10 wt\%)  in water/PG (995 $\mu$L, $x_{\rm H_2O}$ = 0.85) to obtain 0.05 wt\% suspensions.

\subsubsection*{Experimental Design}

Unless stated otherwise, each experiment was performed by depositing a $0.5 \, {\rm \mu L}$ binary droplet of water and propylene glycol (with or without additives) on a clean glass slide placed on a homemade inverted microscope with the possibility of switching between bright-field and dark-field illumination and equipped with a CMOS camera for monochrome (Thorlabs, DCC1545M) or RGB (Thorlabs, DCC1645C) imaging. A custom-made environmental chamber (Okolab) enclosed the microscope to shield the system from external air flows as well as to control temperature ($T = 21 \pm 0.5 \, \degree {\rm C}$) and relative humidity ($RH = 50 \pm 5\%$). The entire setup was mounted on a floated optical table in order to reduce vibrations. 

The vapor source consisted of a glass capillary ($640 \, {\rm \mu m}$ inner radius) terminated with a blunt metal needle of variable size at one end. The needle was fixed to the capillary with a silicone-based sealant. The opposite end contained $20 \, {\rm \mu L}$ of DI water held in place through capillary forces by sealing the open end with wax. Alternatively, when a stronger vapor field was required (Figs. \ref{fig:fig4}, \ref{fig:fig5}, S\ref{fig:figS5} and S\ref{fig:figS6}), the same capillary was used on its own with DI water directly placed at its lower opening. The capillary was mounted on a three-axis micrometric stage, so that it could be accurately positioned in space. The vertical distance of the capillary from the substrate was fixed at $z_{\rm s} = 0.5 \, {\rm mm}$. To determine this distance and the droplet's contact angle $\theta_{\rm c}$, a periscope together with a flip mirror allowed us to switch optical path so as to visualize the droplet's side view instead of its basal plane on the CMOS cameras.

At the start of each experiment, the droplet under study was gently deposited on the clean slide with a pipette using either a disposable low-retention pipette tip (Brand, Z740080) or, for more viscous solutions (Figs. \ref{fig:fig4}, \ref{fig:fig5}, S\ref{fig:figS5} and S\ref{fig:figS6}), a standard pipette tip (Eppendorf). Droplets were carefully positioned at the desired initial distance from the source of water vapor using the microscope stage, which was controlled with motorized actuators (Thorlabs, Z812B actuators with $29 \, {\rm nm}$ minimum step) to guarantee the possibility of translating the droplet with respect to the source in all planar directions. The droplet's dynamics were then recorded with low magnification with the CMOS cameras either at $1\, {\rm fps}$ (frames per second) for the droplet tracking experiment or at $10\, {\rm fps}$ elsewhere. In Fig. \ref{fig:fig3}, real-time tracking of droplet motion over large distances was performed by inputting the direct stream from the camera into a custom \textsc{Matlab} script, which located the droplet's center and automatically moved the motorized stage continuously so as to maintain its relative distance $x_{\rm s}$ from the source at the initial set value. This displacement of the stage as a function of time provided the droplet's velocity. The deposition experiments in Figs. \ref{fig:fig4}A to E and S\ref{fig:figS5} were also performed in a similar way, i.e. by moving the stage to keep the distance between droplet and source constant. All droplets were guided by holding the water vapor source at their leading edge. In Figs. \ref{fig:fig4}F, \ref{fig:fig5}A and S\ref{fig:figS6}, the stage was moved in a straight line or according to a preprogrammed pattern at a constant speed. In Figs. \ref{fig:fig4}F and S\ref{fig:figS6}, PEG crystallisation in the supersaturated deposit was manually nucleated with a metallic needle. In Fig. \ref{fig:fig5}B and C, the combination of two droplets was performed by placing one droplet under a fixed vapor source ($R_{\rm s} = 640 \, {\rm \mu m}$) and a second droplet under a second identical source approximately $10 \, {\rm mm}$ away, which could be moved relative to the first. By reducing the inter-source distance to $3.5 \, {\rm mm}$, the two droplets were made approach gradually until coalescence.

\subsubsection*{Profilometry}

The height profiles of the deposits in Figs. \ref{fig:fig4}B and S\ref{fig:figS6}B were measured on a DektakXT Surface Profilometer (Bruker) with a $5 \, {\rm \mu m}$ stylus at $5 \, {\rm mg}$ force. In Fig. \ref{fig:fig4}B, each profile is the average of 5 different profiles, which were acquired 4 $\mu$m apart using a scan resolution of $0.5 \, {\rm \mu m/pt}$, then levelled and smoothed with a Hampel filter (window size: 100 data points) to remove large random spikes. The 2D profiles in Fig. S\ref{fig:figS6}B were obtained by acquiring 1D profiles (5 $\mu$m apart in the direction perpendicular to the scan direction) with the same parameters. Data points were levelled and then smoothed using a Gaussian-weighted moving average (window size: 30 data points along the scan direction and 4 in the perpendicular direction).

\subsubsection*{Particle Tracking Velocimetry (PTV)}

The flows within the droplets in Fig. S\ref{fig:figS7} were visualized by tracking $5 \, {\rm \mu m}$ polystyrene microparticles using a custom \textsc{Matlab} script. At every instant, the position of these microparticles was determined by applying a sequence of three filters to the raw images. First, a top-hat filter with a 3 pixel radius disk structuring element was used to obtain a uniform background. Then, each pixel was renormalized to the 3x3 pixel window around it in order to set the centroid corresponding to each particle to a value of one. Finally, particle positions were extracted by applying a binary threshold to the processed images after excluding false positives. Trajectories were obtained from these data using a nearest neighbor algorithm between consecutive frames.

\subsection*{Supplementary Text}

\subsubsection*{Local Vapor Pressure at the Droplet's Free Surface}

The steady-state Poisson diffusion equation can be solved for a disc source of radius $R_{\rm s}$ to obtain the vapor pressure of water, $p_{\rm H_2O}(r,\theta,h(r))$, at the droplet's free surface in the presence of a substrate (Fig. S\ref{fig:figS1}A). Assuming to a first approximation that the influence of the droplet on the local vapor pressure is negligible with respect to the source, the solution to the diffusion problem is given by \cite{Warrick}
\begin{equation}\label{eqS1}
 	P_{\rm H_2O} =  \frac{2(p_{\rm s} - p_{RH})}{\pi} \sin^{-1}\bigg( \frac{2R_{\rm s}}{\sqrt{(d-R_{\rm s})^2 + (z_{\rm s} - h)^2}  +  \sqrt{(d+R_{\rm s})^2 + (z_{\rm s} - h)^2}} \bigg)
\end{equation}
where $d(r,\theta) =  \sqrt{ (x_{\rm s} - r\cos{\theta})^2  + (y_{\rm s} - r\sin{\theta})^2}$, $h(r) = \frac{(R_{\rm D}^2-r^2)}{2R_{\rm D}}\theta_{\rm c}$ \cite{Kim2018}, $p_{\rm s}$ is the saturated partial pressure of water at the source (e.g. $p_{\rm s} = 2.49 \, {\rm kPa}$ at $T = 21 \, \degree {\rm C}$ \cite{Lide2004}), and $p_{RH}$ is the partial pressure of water corresponding to the ambient relative humidity $RH$. To account for the presence of the substrate, $p_{\rm H_2O}$ is then calculated by adding to $P_{\rm H_2O}$ the vapor pressure $P'_{\rm H_2O}$ generated by the specular image of the disc source with respect to the surface \cite{Crank1975}, i.e. centered in ($x_{\rm s}, y_{\rm s},-z_{\rm s}$), as 
\begin{equation}\label{eqS2}
	p_{\rm H_2O}(r,\theta,h) = P_{\rm H_2O}(r,\theta,h) + P'_{\rm H_2O}(r,\theta,h)+ p_{RH}. 
\end{equation}

\subsubsection*{Local Surface Tension at the Droplet's Free Surface}

At every instant $t$, due to the preferential evaporation of its more volatile component (water), we can assume that the bulk composition of the droplet $x^{\rm b}_{\rm H_2O}(t)$ changes to a new value $x^{\tau}_{\rm H_2O}(t) = x^{\rm b}_{\rm H_2O}(t + \tau)$ after a sufficiently small time $\tau$ has elapsed. During this change, we can estimate an equivalent mole fraction of water $\chi_{\rm H_2O}(r,\theta,h)$ on the droplet's free surface with respect to the new reference bulk value $x^{\tau}_{\rm H_2O}(t)$ from $p_{\rm H_2O}(r,\theta,h)$. At a given instant $t$, $\chi_{\rm H_2O}(r,\theta,h)$ can then be calculated, using the definition of mole fraction and assuming negligible variations in droplet's volume, as
\begin{equation}\label{eqS3}
	\chi_{\rm H_2O}(r,\theta,h) = \frac{x^{\tau}_{\rm H_2O}V^{-1}_{\rm m}(x^{\tau}_{\rm H_2O})+k p_{\rm H_2O}}{V^{-1}_{\rm m}(x^{\tau}_{\rm H_2O})+k p_{\rm H_2O}} 
\end{equation}
where $V_{\rm m}(x^{\tau}_{\rm H_2O})$ is the molar volume of the mixture at $x^{\tau}_{\rm H_2O}$ as estimated from empirical formulae \cite{Khattab2017} and $k = 0.31 \, {\rm mol Pa^{-1} m^{-3}}$ is a constant (independent of mole fraction, vapor pressure and time) obtained from fitting our model to all experimental data in Fig.~\ref{fig:fig3}. Numerically, at every time step, $x^{\tau}_{\rm H_2O}$ was estimated from the experimentally determined volume of the droplet at the following time step (see Supplementary Text, Time Dependence of Droplet's Geometry and Composition). Interestingly, the meaning of $k p_{\rm H_2O}$ is that of an effective local reduction in the amount of water per volume that leaves the droplet's free surface at any point, and this reduction is proportional to the local vapor pressure of water. Finally, given the local composition $\chi_{\rm H_2O}(r,\theta,h)$ along the droplet's free surface, the local surface tension $\gamma(r,\theta,h)$ can be estimated using empirical formulae \cite{Hoke1992}. The reference value of surface tension $\gamma_{RH}$ used in Fig. \ref{fig:fig2}A is calculated in a similar way after removing the influence of the source in Eq. 5, i.e. by imposing $p_{\rm H_2O} = p_{RH}$.

\subsubsection*{Force Exerted on the Droplet by the Vapor Source}

Following Brochard \cite{Brochard1989}, a droplet can be driven into motion on a horizontal surface because of two reasons: first, the surface tension $\gamma$ at the liquid-air interface is not uniform, thus giving rise to Marangoni flows; second, the spreading coefficient, $S(\gamma) = \Delta \gamma_{\rm s} - \gamma$, depends on position, thus generating a potential landscape for the droplet. Here, $\Delta \gamma_{\rm s} = \gamma_{\rm sv} - \gamma_{\rm sl}$ captures the contribution of the solid surface on $S$ and is the balance between the surface tension of the solid-air interface ($\gamma_{\rm sv}$) and that of the solid-liquid interface ($\gamma_{\rm sl}$). In our case, $S$ varies because of $\gamma$, while, to first approximation, we can assume that the vapor field generated by the source does not alter the contribution from the solid $\Delta \gamma_{\rm s}$.

At any time $t$, the total force acting on the droplet can therefore be estimated by balancing three contributions: a driving force ${\bf F}^{\rm d}$ due to the non-homogenous $S$ and $\gamma$, a dissipative force ${\bf F}^{\rm \gamma}$ associated to the gradient in $\gamma$ and a viscous force ${\bf F}^{\rm v}$ that opposes the previous two contributions so that ${\bf F}^{\rm v} = {\bf F}^{\rm d} + {\bf F}^{\rm \gamma}$. We shall evaluate the three contributions along the direction of motion $x$ (Fig. S\ref{fig:figS1}A) independently in what follows. Due to symmetry in the problem, the balance of forces along the $y$ direction is null.

For small surface tension gradients, $F^{\rm d}_x$ can be calculated by solving the following path integral along the contact line in polar coordinates \cite{Brochard1989}:
 \begin{equation}\label{eqS4}
	F^{\rm d}_x = \int_{0}^{2\pi} \big( R_{\rm D} \frac{\partial (S+\gamma)}{\partial r} \cos \theta - \frac{\partial (S+\gamma)}{\partial \theta} \sin \theta \big) \, R_{\rm D} d\theta.
\end{equation}
As the water vapor does not appreciably change the contribution from the solid in our configuration (i.e. $\Delta \gamma_{\rm s}$ is constant), the solution to the previous integral is straightforward
 \begin{equation}\label{eqS5}
	F^{\rm d}_x = R_{\rm D} \int_{0}^{2\pi} \big( R_{\rm D} \frac{\partial \Delta \gamma_{\rm s}}{\partial r} \cos\theta - \frac{\partial \Delta \gamma_{\rm s}}{\partial \theta} \sin\theta \big) \, d\theta = 0.
\end{equation}

The two remaining force terms can be then calculated by evaluating the integral of the viscous stress at the liquid-solid interface to calculate the total viscous force \cite{Brochard1989}. The functional form of this viscous stress is determined by the flows within the droplet during its motion. For small contact angles $\theta_{\rm c}$, the flow patterns in the droplet advancing with a constant velocity $v_x$ can be calculated using the lubrication approximation, and they are the superposition of a Poiseuille flow induced by the pressure gradient and the Marangoni flows due to the gradient in surface tension at the liquid-air interface \cite{Brochard1989}. As dissipation is controlled by the wedge where the shear gradient is the sharpest \cite{Brochard1989}, the resulting dissipative force, $F^{\rm v}_x - F^{\rm \gamma}_x$, along $x$ can also be calculated by evaluating a path integral along the contact line in polar coordinates, namely
\begin{equation}\label{eqS6}
 	F^{\gamma}_x = \frac{1}{2} \int_{0}^{2\pi} \big( R_{\rm D} \frac{\partial \gamma}{\partial r} \cos \theta - \frac{\partial \gamma}{\partial \theta} \sin \theta \big) \, R_{\rm D} d\theta = \frac{R_{\rm D}}{2} \int_{0}^{2\pi} \big( R_{\rm D} \frac{\partial \gamma}{\partial r} \cos \theta - \frac{\partial \gamma}{\partial \theta} \sin \theta \big) \, d\theta
\end{equation}
and
\begin{equation}\label{eqS7}
 	F^{\rm v}_x = \int_{0}^{2\pi} \frac{3\eta\ell_{\rm n}v_x}{\theta_{\rm c}}  \, R_{\rm D} d\theta = \frac{6\pi\eta R_{\rm D} \ell_{\rm n}v_x}{\theta_{\rm c}}
\end{equation}
where $\eta$ is the dynamic viscosity of the mixture determined empirically at different compositions \cite{Khattab2017} and $\ell_{\rm n} = 11.2$ is a cutoff constant \cite{Brochard1989,Cira2015}. Interestingly, as $F^{\rm d}_x = 0$, we can see how our process is entirely driven and dominated by dissipative phenomena.

Combining all force contributions, the velocity of the droplet in the vapor field generated by the source is then given at any instant $t$ by 
\begin{equation}\label{eqS8}
 	v_x = \frac{\theta_{\rm c}}{12\pi\eta \ell_{\rm n}} \int_{0}^{2\pi} \big( R_{\rm D} \frac{\partial \gamma}{\partial r} \cos \theta - \frac{\partial \gamma}{\partial \theta} \sin \theta \big) \, d\theta.
\end{equation}

\subsubsection*{Time Dependence of Droplet's Geometry and Composition}

The geometry of the droplet over time was estimated experimentally from videos of $0.5 \, {\rm \mu L}$ droplets of initial composition $x_{\rm H_2O} = 0.95$ evaporating in the presence of the source (Fig. S\ref{fig:figS1}). Radius $R_{\rm D}$ and contact angle $\theta_{c}$ were measured directly in intervals of 25 s (Fig. S\ref{fig:figS1}B and C). Assuming a spherical cap geometry \cite{Kim2018}, these values were then used to estimate the droplet's volume $V_{\rm D}$ as well as to define the droplet's free surface in cylindrical coordinates in time (Fig. S\ref{fig:figS1}A and D). Each set of data was fitted to a continuous function given by 5\textsuperscript{th}-order polynomials. By assuming that no PG evaporates due to its high boiling point at $187.4 \degree \, {\rm C}$ \cite{Schierholtz1935}, all volume loss during evaporation can be attributed to water, so the bulk composition of the droplet, to a first approximation, can be given by:
\begin{equation}\label{eqS9}
	x^{\rm b}_{\rm H_2O}(t) = \frac{M_{\rm PG}\rho_{\rm H_2O}(V_{\rm D}(t) - V^0_{\rm PG})}{M_{\rm PG}\rho_{\rm H_2O}V_{\rm D}(t) + (M_{\rm H_2O}\rho_{\rm PG} - M_{\rm PG}\rho_{\rm H_2O})V^0_{\rm PG}}
\end{equation}
where $M_{\rm PG} = 76.1 \, {\rm g mol^{-1}}$, $M_{\rm H_2O} = 18.0 \, {\rm g mol^{-1}}$, $\rho_{\rm PG} = 1037 \, {\rm kg m^{-3}}$ and $\rho_{\rm H_2O}= 998 \, {\rm kg m^{-3}}$ are respectively the molecular weights and densities of propylene glycol and water at $21  \degree \, {\rm C}$, and $V^0_{\rm PG} = 0.09 \, {\rm \mu L}$ is the volume of propylene glycol, which can be estimated from Eq. \ref{eqS9} at $t=0$ as the initial mole fraction of water is known. The values of $x^{\rm b}_{\rm H_2O}$ in time were then used to estimate the time variation of composition-dependent physical parameters, such as the equivalent mole fraction of water on the droplet's free surface (Eq. \ref{eqS3}) and the dynamic viscosity $\eta$ of the mixture (Eq. \ref{eqS8}), so that the time dependence of surface tension, forces and droplet's velocity could be evaluated numerically.

\clearpage

\clearpage

\subsection*{Supplementary Figures}

\makeatletter
\renewcommand{\fnum@figure}{\figurename~S\thefigure}
\makeatother
\setcounter{figure}{0}

\begin{figure}[h]
\centering
\includegraphics[width=\textwidth]{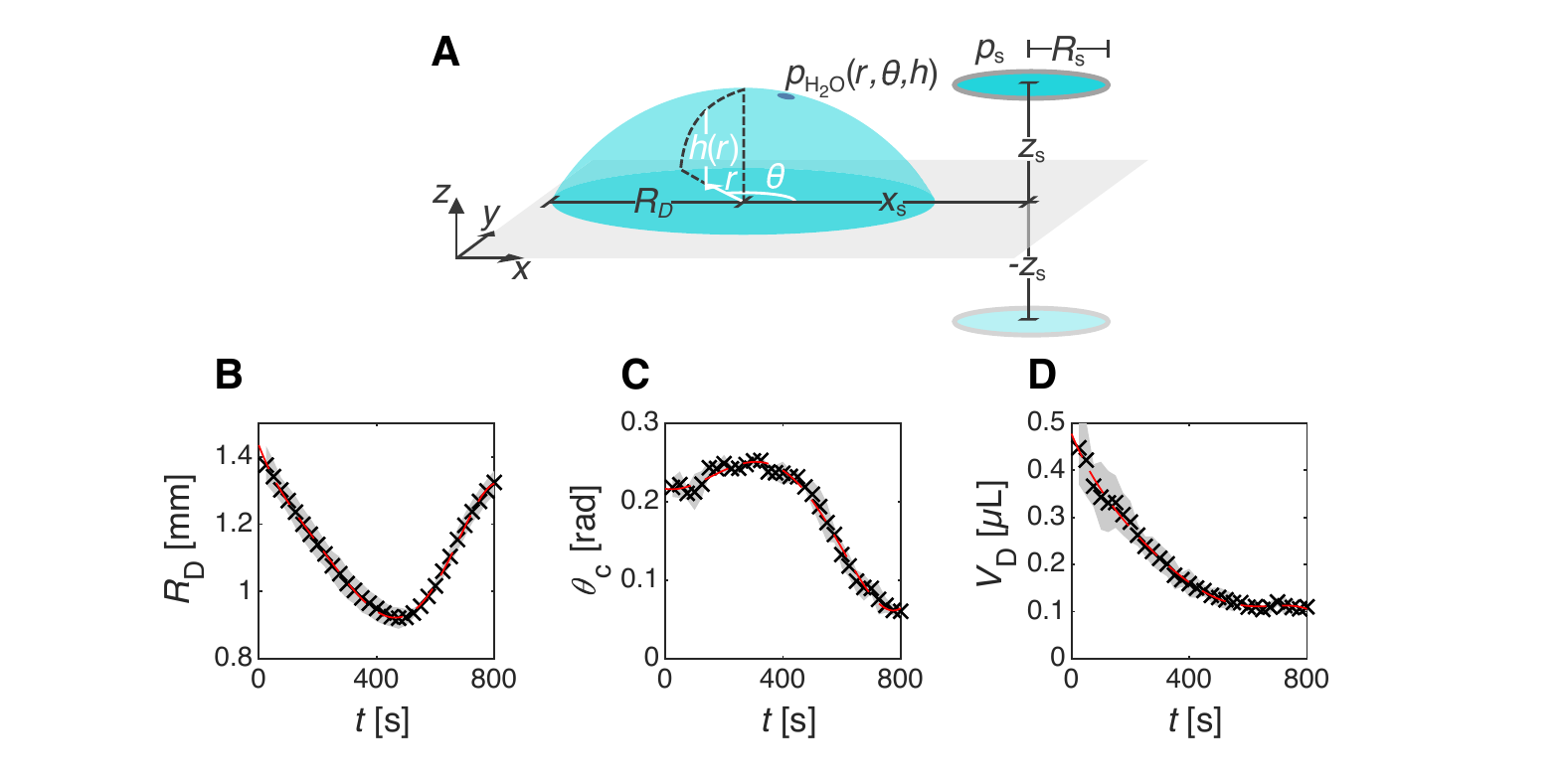}
\caption{{\bf Time evolution of droplet's geometry.} ({\bf A}) Schematic geometry of the droplet (radius $R_{\rm D}$) and the vapor source (radius $R_{\rm s}$, saturated vapor pressure $p_{\rm s}$) used in the model. The droplet's free surface is defined in cylindrical coordinates $(r,\theta,h)$ and a local water vapor pressure, $p_{\rm H_2O}(r,\theta,h)$, is associated to each of its points; $x_{\rm s}$ is the in-plane distance between droplet and source and $z_{\rm s}$ is the vertical distance between source and surface. The specular image of the source with respect to the surface is also shown at $-z_{\rm s}$. ({\bf B-D}) Mean experimental values (crosses) of droplet's ({\bf B}) radius $R_{\rm D}$, ({\bf C}) contact angle $\theta_{\rm c}$ and ({\bf D}) volume $V_{\rm D}$ over time. The mean values are averages over 3 different droplets evaporating in the presence of the source. The droplet's volume $V_{\rm D}$ is calculated from the values of $R_{\rm D}$ and $\theta_{c}$ assuming a spherical cap geometry. The shaded areas represent one standard deviation around the mean values. Each variable is fitted to a 5\textsuperscript{th}-order polynomial (dashed line).}
\label{fig:figS1}
\end{figure}

\clearpage

\begin{figure}[h]
\centering
\includegraphics[width=\textwidth]{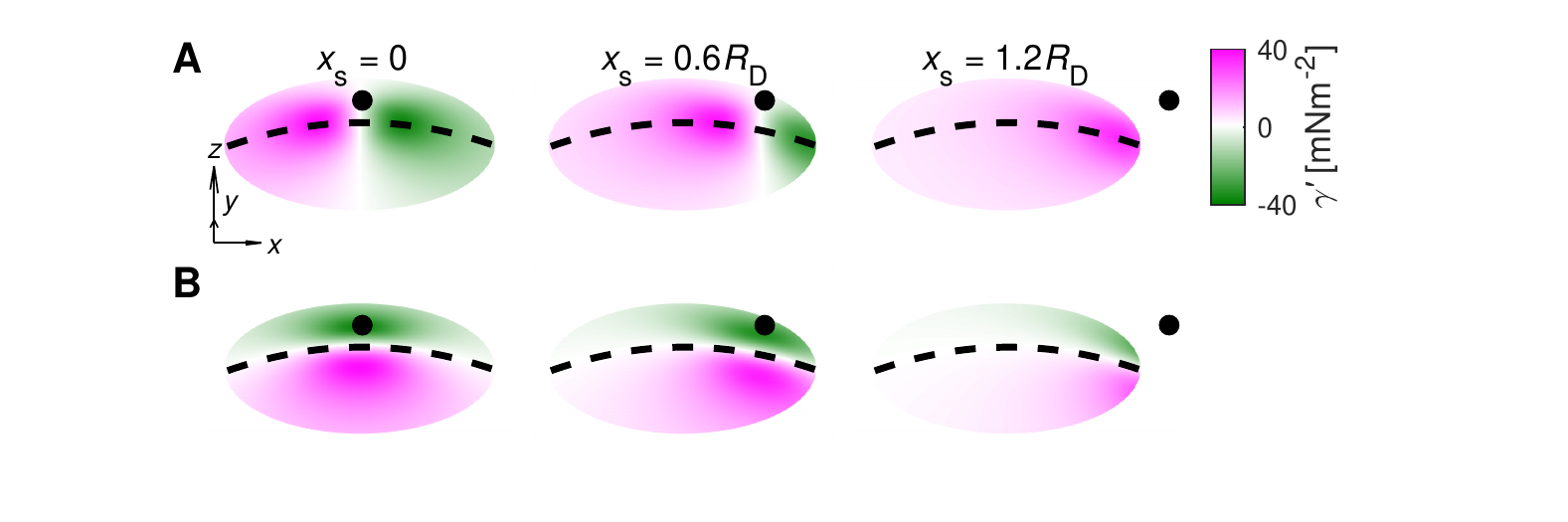}
\caption{{\bf Calculated gradients of surface tension along the droplet's free surface.} ({\bf A-B}) Calculated gradients of surface tension along ({\bf A}) the direction of motion ($x$) and ({\bf B}) the in-plane perpendicular direction ($y$) due to vapor sources placed at different positions $x_{\rm s}$ (black dots, $R_{\rm s} = 350 \, {\rm \mu m}$) when evaporation starts. These gradients correspond to the profiles of surface tension shown in Fig. \ref{fig:fig2}A. Dashed lines: meridians through the droplets' apices. The coordinate unit vectors correspond to 0.5 mm.}
\label{fig:figS2}
\end{figure}

\clearpage

\begin{figure}[h]
\centering
\includegraphics[width=\textwidth]{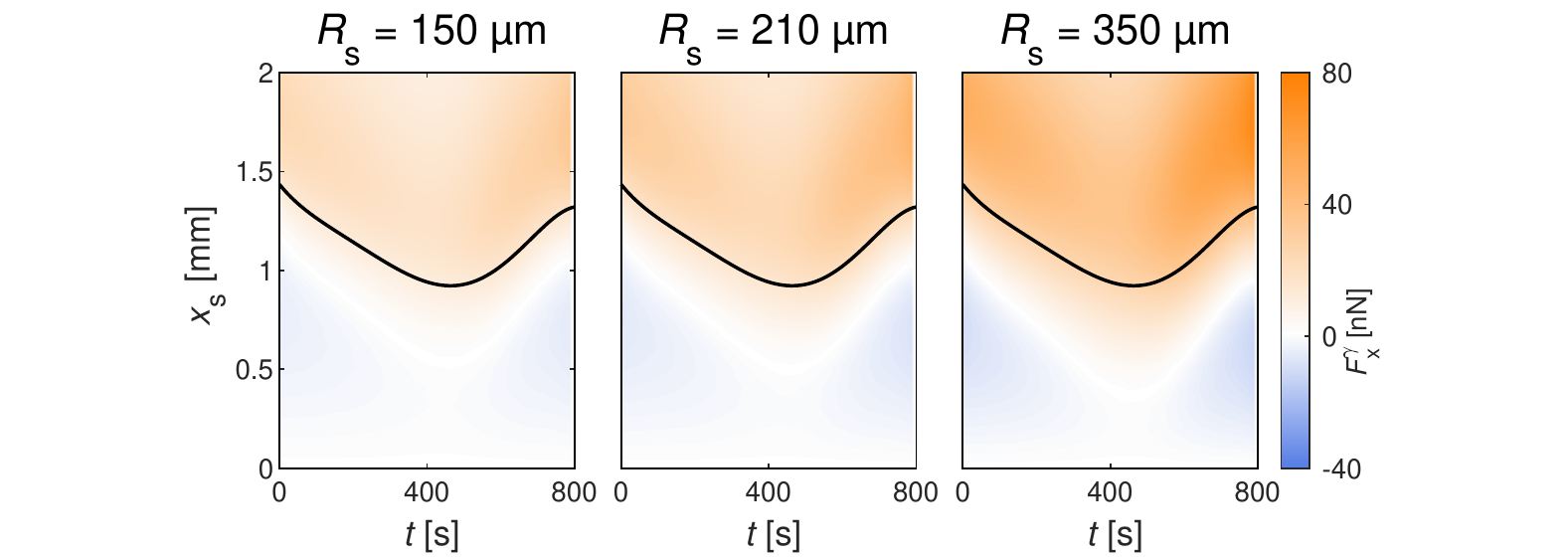}
\caption{{\bf Calculated droplet's driving force.} Calculated driving force $F^{\rm \gamma}_x$  (Eq. \ref{eq1}) exerted on the droplet by the vapor source as a function of the distance $x_{\rm s}$ from the source and time $t$ from the beginning of the evaporation for increasing source radii $R_{\rm s}$. The solid lines represent the time evolution of the droplet's radius $R_{\rm D}$ (Fig. S\ref{fig:figS1}B).}
\label{fig:figS3}
\end{figure}

\clearpage

\begin{figure}[h]
\centering
\includegraphics[width=\textwidth]{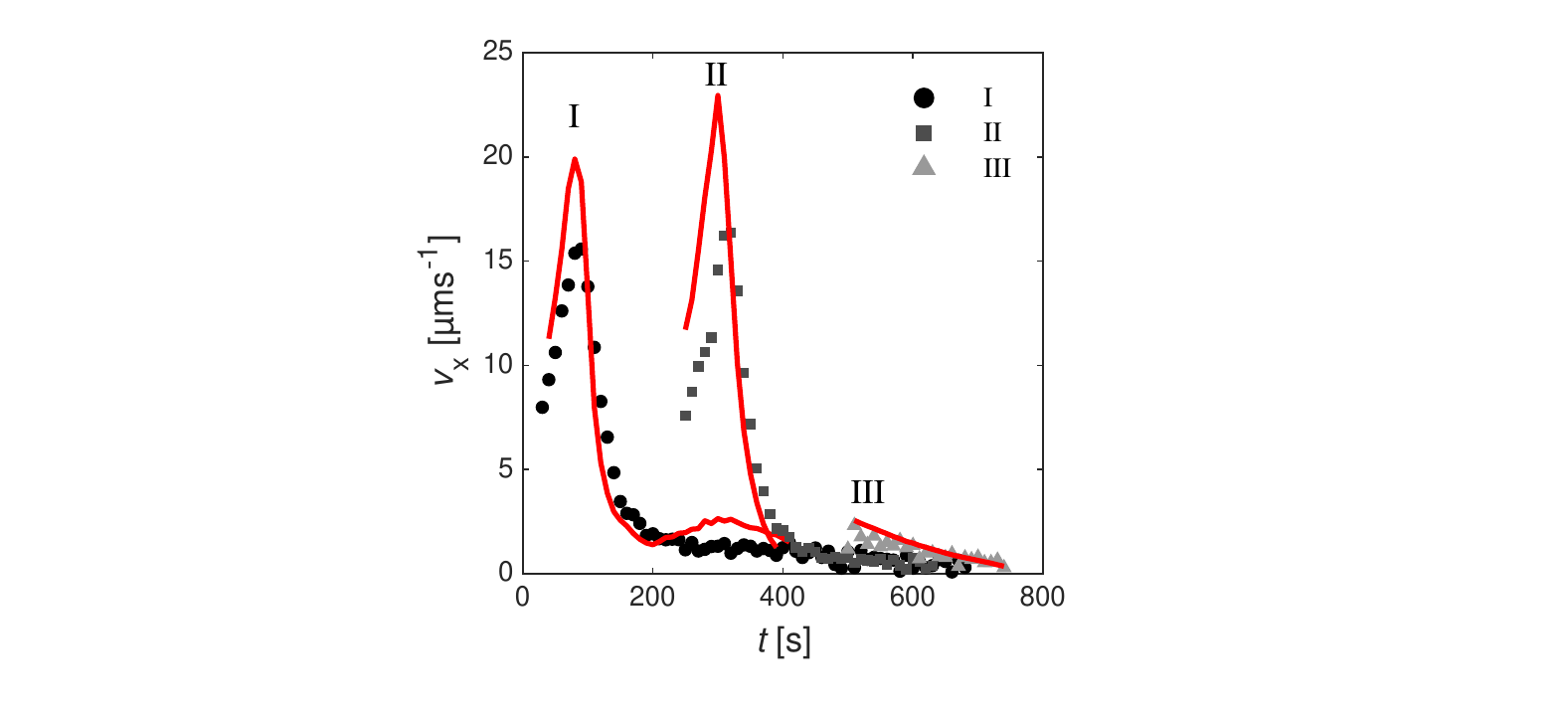}
\caption{ {\bf Velocity of individual droplets moving freely towards the vapor source.}  Time evolution of the velocity (symbols) of three distinct droplets ($V_{\rm D} = 0.5 \, {\rm \mu L}$, $x_{\rm H_2O} = 0.95$) starting to move towards the vapor source ($R_{\rm s} = 350 \, {\rm \mu m}$) from $x_{\rm s} = 2 \, {\rm mm}$ (as in Fig. \ref{fig:fig1}) at (I) $30 \, {\rm s}$, (II) $250 \, {\rm s}$ and (III) $500 \, {\rm s}$ from the beginning of their evaporation. The trajectories for the three droplets are also shown in Fig. \ref{fig:fig3}. The solid lines show the predicted velocity for a given droplet as interpolated from the experimental data for $R_{\rm s} = 350 \, {\rm \mu m}$ in Fig. \ref{fig:fig3}. Beyond an overestimation of the peak velocity, these interpolated data predict the functional form of the droplet's velocity well.}
\label{fig:figS4}
\end{figure}

\clearpage

\begin{figure}[h]
\centering
\includegraphics[width=\textwidth]{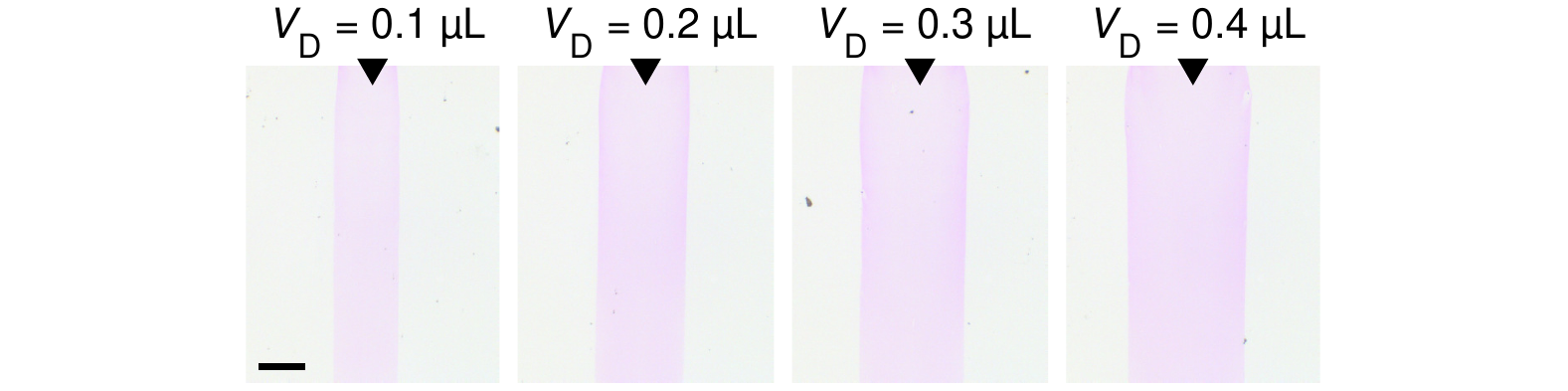}
\caption{{\bf Printing with moving droplets: influence of volume at a fixed PVOH concentration.} Photographs of deposits from moving water/PG droplets ($x_{\rm H_2O}$ = 0.95) of same PVOH concentration ($[{\rm PVOH}] = 2 {\rm mM}$) and increasing volumes. All droplets were guided along a line by holding a water vapor source ($R_{\rm s}$ = 640 $\mu$m) at their leading edge and contain rhodamine B for ease of visualization. In the photographs, the background was subtracted and the black triangles indicate the direction of motion. Scale bar: 1 mm.}
\label{fig:figS5}
\end{figure}

\clearpage

\begin{figure}[h]
\centering
\includegraphics[width=\textwidth]{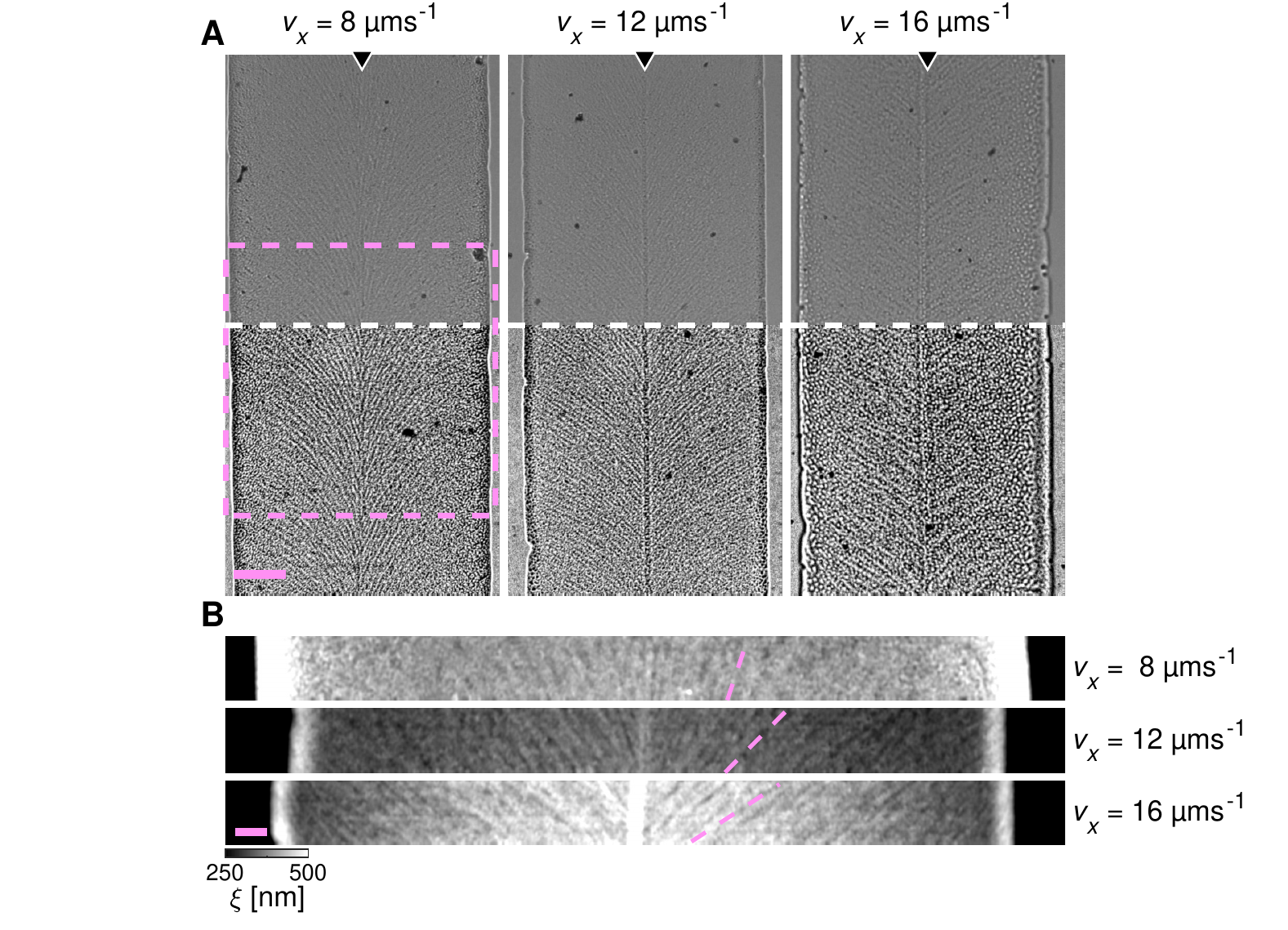}
\caption{{\bf Alignment of PEG deposits with moving droplets.} ({\bf A}) Photographs of alignment in linear polymer deposits ($[{\rm PEG}] = 100 \, {\rm mM}$) from moving water/PG droplets ($V_{\rm D} = 0.5 \, {\rm \mu L}$, $x_{\rm H_2O}$ = 0.95) guided at different speeds. The dashed lines separate two parts in the images where the top half is taken from the original photographs and the bottom half is enhanced with an edge-aware filter for contrast. The part delimited by a dashed box in the image at $8 \, {\rm \mu ms^{-1}}$ corresponds to Fig. \ref{fig:fig4}F. All droplets were guided at a constant velocity by a water vapor source ($R_{\rm s} = 640 \, {\rm \mu m}$). The black triangles indicate the direction of motion. Scale bar: 0.5 mm. ({\bf B}) Examples of bidimensional height maps for the polymer deposits in {\bf A} \cite{Supplementary}. The dashed lines highlight the directionality of the ridges in the deposit. Scale bar: $100 \, {\rm \mu m}$.}
\label{fig:figS6}
\end{figure}
\clearpage

\clearpage

\begin{figure}[h]
\centering
\includegraphics[width=\textwidth]{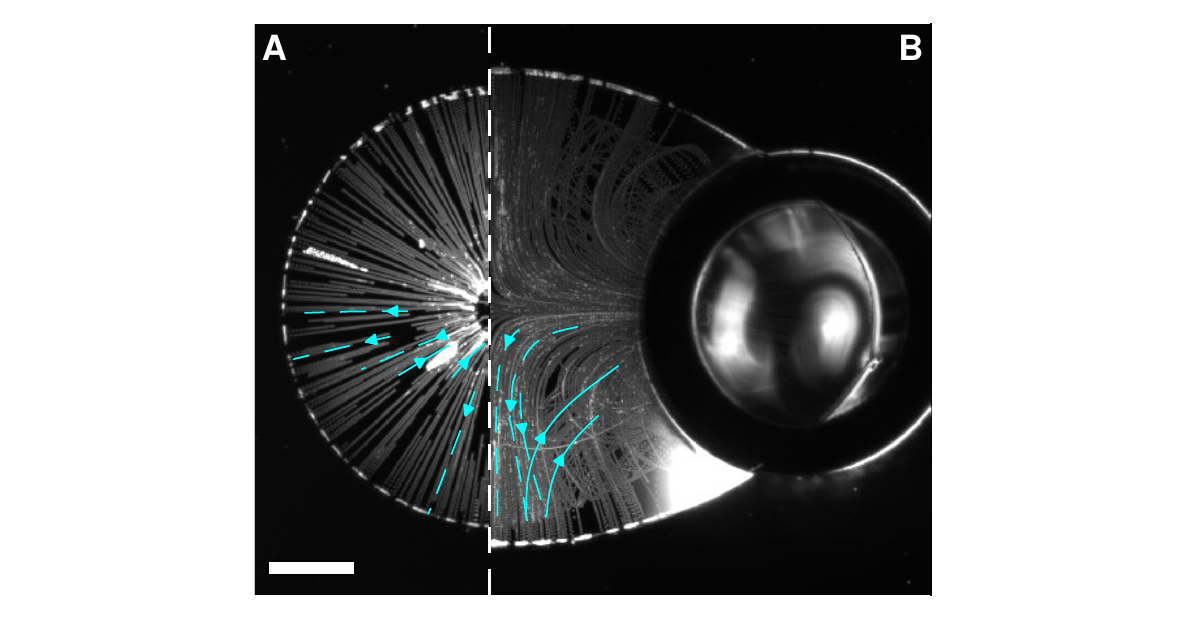}
\caption{{\bf Flows within two coalesced droplets}. Flows ({\bf A}) in a water/PG droplet ($V_{\rm D} = 0.5 \, {\rm \mu L}$, $x_{\rm H_2O}$ = 0.85) in the absence of a water vapor source and ({\bf B}) in a coalesced droplet formed by two similar droplets held by two vapor sources ($R_{\rm s} = 640 \, {\rm \mu m}$) as in Figs. \ref{fig:fig5}B and C \cite{Supplementary}. The two cases are shown side by side and are separated by a vertical white dashed line. The cyan triangles show the directionality of typical flow lines near the droplet's basal plane (dashed lines) and along its free surface (solid lines). The flows were measured by PTV over $30 \, {\rm s}$ \cite{Supplementary}. Scale bar: $0.5 \, {\rm mm}$.}
\label{fig:figS7}
\end{figure}

\end{document}